\documentclass[twocolumn,english,superscriptaddress, prl]{revtex4-1}
\usepackage[T1]{fontenc}
\usepackage[latin9]{inputenc}
\usepackage{color}
\usepackage{units}
\usepackage{mathrsfs}
\usepackage{amsmath}
\usepackage{amssymb}
\usepackage{graphicx}

\makeatletter
\usepackage[caption=false]{subfig}

\@ifundefined{showcaptionsetup}{}{%
	\PassOptionsToPackage{caption=false}{subfig}}
\usepackage{subfig}
\makeatother

\usepackage{babel}
\begin{document}
	
	\title{Quantum phase-sensitive diffraction and imaging
	using entangled photons}
	
	\author{Shahaf Asban}
	\email{sasban@uci.edu}

	\affiliation{Department of Chemistry and Physics and Astronomy, University of
		California, Irvine, California 92697-2025, USA}
	
	\author{Konstantin E. Dorfman }
	\email{dorfmank@lps.ecnu.edu.cn}

	\affiliation{State Key Laboratory of Precision Spectroscopy, East China Normal
		University, Shanghai 200062, China}
	
	\author{Shaul Mukamel}
	\email{smukamel@uci.edu}

	\affiliation{Department of Chemistry and Physics and Astronomy, University of
		California, Irvine, California 92697-2025, USA}

\begin{abstract}
We propose a novel quantum diffraction imaging technique whereby one
photon of an entangled pair is diffracted off a sample and detected
in coincidence with its twin. The image is obtained by scanning the
photon that did not interact with matter. We show that when a dynamical
quantum system interacts with an external field, the phase information
is imprinted in the state of the field in a detectable way. The contribution
to the signal from photons that interact with the sample scales as
$\propto I_{p}^{\nicefrac{1}{2}}$, where $I_{p}$ is the source intensity,
compared to $\propto I_{p}$ of classical diffraction. This makes
imaging with weak-field  possible, avoiding damage to delicate samples.
A Schmidt decomposition of the state of the field can be used for
image enhancement by reweighting the Schmidt modes contributions. 
\end{abstract}

\maketitle
Rapid advances in short-wavelength ultrafast light sources, have revolutionized
our ability to observe the microscopic world. With bright Free Electron
Lasers and high harmonics tabletop sources, short time (femtosecond)
and length (subnanometer) scales become accessible experimentally.
These offer new exciting possibilities to study spatio-spectral properties
of quantum systems driven out of equilibrium, and monitor dynamical
relaxation processes and chemical reactions. The spatial features
of small-scale charge distributions can be recorded in time. Far-field
off-resonant X-ray diffraction measurements provide useful information
on the charge density $\sigma\left(\boldsymbol{Q}\right)$ where $\boldsymbol{Q}$
is the diffraction wavector. The observed diffraction pattern $S\left(\boldsymbol{Q}\right)$
is given by the modulus square $S\left(\boldsymbol{Q}\right)\propto\left|\sigma\left(\boldsymbol{Q}\right)\right|^{2}$.
Inverting these signals to real-space $\sigma\left(\boldsymbol{r}\right)$ requires a Fourier transform.
Since the phase of $\sigma\left(\boldsymbol{Q}\right)$ is not available,
the inversion requires phase retrieval which can be done using either
algorithmic solutions \cite{Elser:03,Fienup1982} or more sophisticated and costly
experimental setups such as heterodyne measurements \cite{Marx2008}.
Correlated beam techniques \cite{Walborn2010,Erkmen2010,Gatti2004,Howell2004,Edgar2012,Aspden2013,Lemos2014}
in the visible regime, have been shown to circumvent this problem
by producing the real-space image of mesoscopic objects. Such techniques
have classical analogues
using correlated light, and reveal the modulus square of the studied
object $\left|\sigma\left(\boldsymbol{r}\right)\right|^{2}$ \cite{Shapiro2008,Boyd2008}. 

\begin{figure}[h]
\begin{centering}
\includegraphics[scale=0.37]{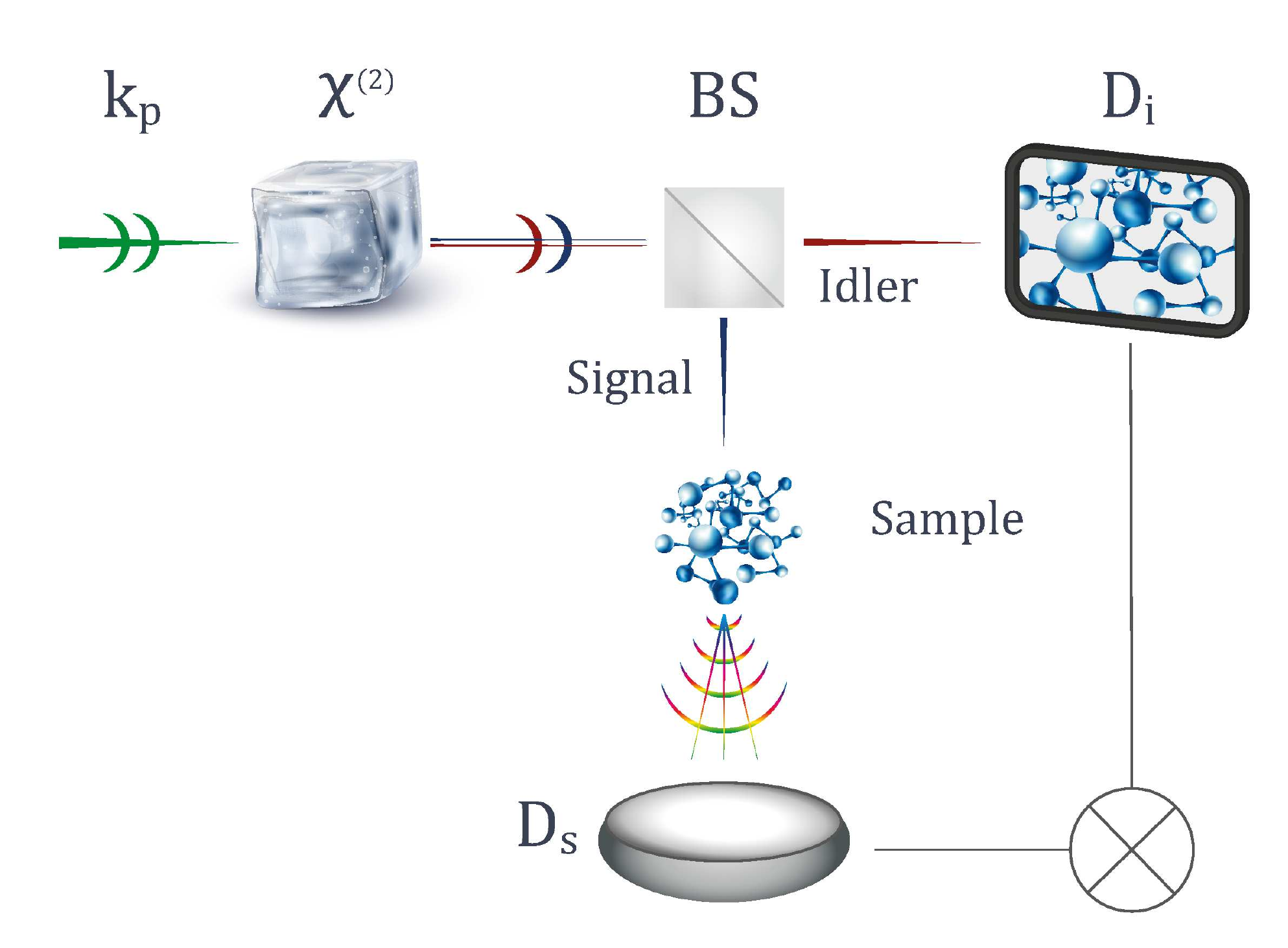}
\par\end{centering}
\caption{\label{The setup} \emph{Sketch of the proposed quantum imaging setup}.
A broad-band pump $\boldsymbol{k}_{p}$ propagates through a $\chi^{\left(2\right)}$
crystal, generating an entangled photon pair denoted as \emph{signal
}and \emph{idler}. The photons are distinguished either by polarization
(type-II) or frequency (type-I) and are separated by a beam-splitter
(BS). The signal\emph{ }photon interacts with the sample, and can
be further frequency dispersed and collected by a 'bucket' detector
$D_{s}$ with no spatial resolution. The idler\emph{ }is spatially
resolved in the transverse plane by the detector $D_{i}$. The two
photons are detected in coincidence as defined in Eq.$\left(\text{\ref{eq:  intensity-intensity definition}}\right)$. }
\end{figure}

In this paper we consider the setup shown in Fig.$\left(\text{\ref{The setup}}\right)$.
We focus on off-resonant scattering of entangled photons in which
only one photon, denoted as the "signal", interacts with a sample. Its entangled
counterpart, the "idler", is spatially scanned and measured in coincidence
with the arrival of the signal photon. The idler reveals the image
and also uncovers \emph{phase information}, as was recently shown in \cite{Dorfman2018}
for linear diffraction. 

Our first main result is that for small diffraction angles, using Schmidt
decomposition of the two-photon amplitude $\Phi\left(\boldsymbol{q}_{s},\boldsymbol{q}_{i}\right)=\sum_{n}^{\infty}\sqrt{\lambda_{n}}u_{n}\left(\boldsymbol{q}_{s}\right)v_{n}\left(\boldsymbol{q}_{i}\right)$
where $\lambda_{n}$ is the respective mode weight - reads,

\noindent 
\begin{equation}
{\cal S}^{\left(p\right)}\left[\bar{\boldsymbol{\rho}}_{i}\right]\propto\mathfrak{Re}\sum_{nm}\sqrt{\lambda_{n}\lambda_{m}}\beta_{nm}^{\left(p\right)}v_{n}^{*}\left(\boldsymbol{\bar{\rho}}_{i}\right)v_{m}\left(\boldsymbol{\bar{\rho}}_{i}\right).\label{eq: small angel main eq}
\end{equation}

\noindent Here  $\beta_{nm}^{\left(1\right)}=\int d\boldsymbol{r\:}u_{n}\left(\boldsymbol{r}\right)\sigma\left(\boldsymbol{r}\right)u_{m}^{*}\left(\boldsymbol{r}\right)$,
 $\beta_{nm}^{\left(2\right)}=\int d\boldsymbol{r\:}u_{n}\left(\boldsymbol{r}\right)\left|\sigma\left(\boldsymbol{r}\right)\right|^{2}u_{m}^{*}\left(\boldsymbol{r}\right)$ and $\boldsymbol{\bar{\rho}}_{i}$ represents the transverse detection plane.
 $\sigma\left(\boldsymbol{r}\right)$ is the charge
density of the target object prepared by an actinic pulse and $p=\left(1,2\right)$ represents
the order in $\sigma\left(\boldsymbol{r}\right)$.
For large diffraction angels and frequency-resolved signal, the phase
dependent image is modified to ${\cal S}\left[\bar{\boldsymbol{\rho}}_{i}\right]\propto\mathfrak{Re}\sum_{nm}^{\infty}\gamma_{nm}\sqrt{\lambda_{n}\lambda_{m}}v_{n}^{*}\left(\boldsymbol{\bar{\rho}}_{i}\right)v_{m}\left(\boldsymbol{\bar{\rho}}_{i}\right)$,
where $\gamma_{nm}$ have a similar structure to $\beta_{nm}^{\left(1\right)}$
modulated by the Fourier decomposition of the Schmidt basis. $\gamma_{nm}$
is phase dependent in contrast to diffraction with classical
sources.

Our second main result tackles the spatial resolution enhancement.
In entanglement-based imaging, the resolution is limited by the degree
of correlation of the two beams. Schmidt decomposition of the image allows to enhance
desired spatial features of the charge density. High order Schmidt
modes (which correspond to angular momentum transverse modes with
high topological charge) offer more detailed matter information. Reweighting
of Schmidt modes  maximizes modal entropy which yields matter information gain and reveals
fine details of the charge density. Moreover, ${\cal S}^{\left(1\right)}$
in Eq.$\left(\text{\ref{eq: small angel main eq}}\right)$ has no
classical analogue, the contribution to the over-all signal from the
"signal" photons scales as ${I}_{p}^{\nicefrac{1}{2}}$ where ${I}_{p}$
is the intensity of the source. This is a unique signature of the linear diffraction
 \cite{Dorfman2018}. The over-all detected signal is obtained
in coincidence and scales as $\propto {I}_{p}^{\nicefrac{3}{2}}$. Classical
diffraction in contrast requires two interactions with the incoming field and
therefore scales as $I_{p}$, and the corresponding coincidence scales
as $\propto {I}_{p}^{2}$, which also applies for ${\cal S}^{\left(2\right)}$.
Thanks to this favorable scaling, weak fields can be used to study fragile samples
in order to avoid damage.

\section{Spatial entanglement}

Various sources of entangled photons are available, from quantum
dots \cite{Huber2017}, to cold atomic gasses \cite{Liang2018} and
nonlinear crystals which are reviewed in \cite{Walborn2010}. A general
two-photon state can be written in the form,

\begin{equation}
\vert\psi\rangle=\sum_{\boldsymbol{k}_{s},\boldsymbol{k}_{i}}\Phi\left(\boldsymbol{k}_{s},\boldsymbol{k}_{i}\right)\epsilon_{\boldsymbol{k}_{s}}^{\left(\mu_{s}\right)}\epsilon_{\boldsymbol{k}_{i}}^{\left(\mu_{i}\right)}a_{\boldsymbol{k}_{s},\mu_{s}}^{\dagger}a_{\boldsymbol{k}_{i},\mu_{i}}^{\dagger}\vert0_{s},0_{i}\rangle,
\end{equation}

\noindent where $\epsilon_{\boldsymbol{k}}^{\left(\nu\right)}$ is
polarization, $a_{\boldsymbol{k,\nu}}\left(a_{\boldsymbol{k},\nu}^{\dagger}\right)$
are field annihilation (creation) operators and $\Phi\left(\boldsymbol{k}_{s},\boldsymbol{k}_{i}\right)$
is two photon amplitude. In the paraxial approximation the transverse
momentum $\left\{ \boldsymbol{q}_{s},\boldsymbol{q}_{i}\right\} $
and the longitudinal degrees of freedom are factorized. The transverse
amplitude of photon-pair generated using parametric down converter takes then the form, 
\cite{Mandel1985,Walborn2010,Grice1997,Schlawin2013b}, 

\begin{equation}
\Phi\left(\boldsymbol{q}_{s},\boldsymbol{q}_{i}\right)=\Gamma\left(\boldsymbol{q}_{s}+\boldsymbol{q}_{i}\right)\text{sinc}\left(L^{2}\left(\boldsymbol{q}_{s}-\boldsymbol{q}_{i}\right)^{2}\right),\label{eq:Amplitude 2D}
\end{equation}

\noindent here $\Gamma\left(\boldsymbol{q}\right)$ are the pump envelope
of the transverse components, $L^{2}=\nicefrac{l_{z}\lambda_{p}}{4\pi}$
where $\lambda_{p}$ is the central frequency wavelength and $l_{z}$
is the length of the nonlinear crystal along the longitudinal direction.
The state of field  is then given by,

\begin{align}
\vert\psi\rangle & =\vert\text{vac}\rangle+C\sum_{\begin{array}{c}
\boldsymbol{q}_{s},\boldsymbol{q}_{i}\\
\omega_{s},\omega_{i}
\end{array}}A_{p}\left(\omega_{s}+\omega_{i}\right)\Phi\left(\boldsymbol{q}_{s},\boldsymbol{q}_{i}\right)\nonumber \\
 & \times\vert\boldsymbol{q}_{s},\omega_{s};\boldsymbol{q}_{i},\omega_{i}\rangle,\label{wave-function - general term}
\end{align}

\noindent where $C$ is a normalization prefactor and $A_{p}$ is the pump envelope.

\subsection{Schmidt decomposition of entangled two-photon states}

\noindent The hallmark of entangled photon pairs is that they cannot
be considered as two separate entities. This is expressed by the inseparability
of the field amplitude $\Phi$ into a product of single photon amplitude;
all the interesting quantum optical effects discussed below are derivatives
of this feature. $\Phi$ can be represented as a superposition of separable
states using the Schmidt decomposition \cite{Peres1995,Ekert1995,Law2004}.

\begin{equation}
\Phi\left(\boldsymbol{q}_{s},\boldsymbol{q}_{i}\right)=\sum_{n}^{\infty}\sqrt{\lambda_{n}}u_{n}\left(\boldsymbol{q}_{s}\right)v_{n}\left(\boldsymbol{q}_{i}\right),\label{eq:Schmidt decomposition}
\end{equation}

\noindent where the Schmidt modes $u_{n}\left(\boldsymbol{q}_{s}\right)$
and $v_{n}\left(\boldsymbol{q}_{i}\right)$ are the eigenvectors of
the signal and the idler reduced density matrices, and the eigenvalues
$\lambda_{n}$ satisfy the normalization $\sum_{n}\lambda_{n}=1$
\cite{Ekert1995}. The number of relevant modes serves as an indicator
for the degree of inseparability of the amplitude, i.e. photon entanglement.
Common measures for entanglement include the entropy $S_{ent}=-\sum_{n}\lambda_{n}\log_{2}\lambda_{n}$,
or the \emph{Schmidt number} $\kappa^{-1}\equiv\sum_{n}\lambda_{n}^{2}$.
The latter is also known as the \emph{participation ratio }as it quantifies
the number of important Schmidt modes, or the effective joint Hilbert
space size of the two photons. In a maximally entangled wavefunction,
all modes contribute equally.\textcolor{blue}{{} }

\begin{figure}[h]
\begin{centering}
\includegraphics[bb=130bp 255bp 440bp 515bp,clip,scale=0.75]{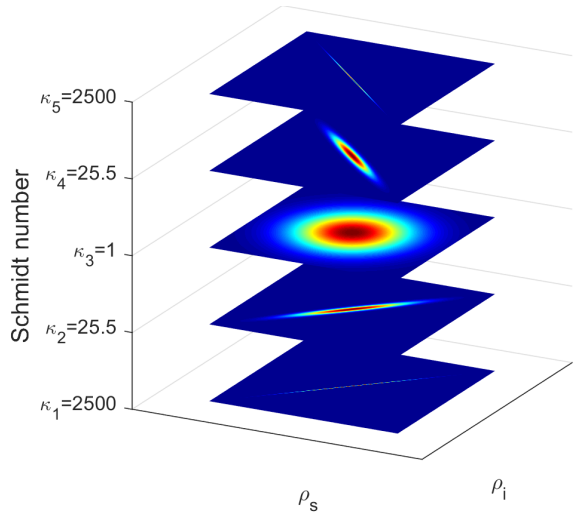}
\par\end{centering}
\raggedright{}\caption{\emph{Transverse beam amplitude profile for different Schmidt Numbers.}
For $\kappa_{3}=1$ the amplitude in Eq.$\left(\text{\ref{eq:Schmidt decomposition}}\right)$
is separable and the photons are not entangled. As $\kappa$ is increased
the amplitude approaches a narrow distribution. $\kappa_{1}=2500$
and $\kappa_{2}=25.5$ are obtained in the $\sigma_{p}L>1$ regime,
the amplitude approaches $\Phi_{\gg}^{\left(\infty\right)}\propto\delta\left(\boldsymbol{q}_{s}+\boldsymbol{q}_{i}\right)$.
$\kappa_{4}=25.5$ and $\kappa_{5}=2500$ are taken in the $\sigma_{p}L<1$
regime, with the asymptotic amplitude $\Phi_{\ll}^{\left(\infty\right)}\propto\delta\left(\boldsymbol{q}_{s}-\boldsymbol{q}_{i}\right)$.\label{fig:Schmidt-numbermVs. Amplitude}}
\end{figure}

The spatial profile of the photons in the transverse plane (perpendicular
to the propagation direction), can be expanded and measured using
a variety of basis functions. E.g. Laguerre-Gauss (LG) or Hermite-Gauss
(HG) have been demonstrated experimentally \cite{Mair2001,Straupe2011}.
These sets satisfy orthonormality $\int d^{2}\boldsymbol{q}\:u_{n}\left(\boldsymbol{q}\right)v_{k}\left(\boldsymbol{q}\right)=\delta_{nk}$
and closure relations $\sum_{n}\:u_{n}\left(\boldsymbol{q}\right)v_{n}\left(\boldsymbol{q}^{'}\right)=\delta^{\left(2\right)}\left(\boldsymbol{q}-\boldsymbol{q}^{'}\right)$.
The deviation of $\lambda_{n}$ from a uniform (flat) distribution
reflects the degree of entanglement. Perfect quantum correlations
correspond to maximal entanglement entropy and thus a flat distribution
of modes. This is further clarified by the closure relations, which
demonstrate the convergence into a point-to-point mapping in the limit
of perfect transverse entanglement. The biphoton amplitude has
two limiting cases for infinite participation ratio which are demonstrated
in Fig.$\left(\ref{fig:Schmidt-numbermVs. Amplitude}\right)$. When
the $\text{sinc}$ function in Eq.$\left(\text{\ref{eq:Amplitude 2D}}\right)$
is approximated by a Gaussian, the Schmidt number is given in a closed
form \cite{Giedke2003}, 

\begin{equation}
\kappa=\frac{1}{4}\left(\sigma_{p}L+\frac{1}{\sigma_{p}L}\right)^{2},\label{eq:Schmidt number}
\end{equation}

\noindent where $\sigma_{p}^{2}$ is the variance of the transverse
momentum of the pump. For $\sigma_{p}=l=1$, we get $\kappa=1$ and
the two-photon wavefunction is separable $\Phi^{\left(\kappa=1\right)}\equiv\Phi^{\left(1\right)}\left(\boldsymbol{q}_{s},\boldsymbol{q}_{i}\right)=\Phi\left(\boldsymbol{q}_{s}\right)\Phi\left(\boldsymbol{q}_{i}\right)$
(no entanglement). A high number of relevant Schmidt modes indicates
stronger quantum correlations between the two photons as shown in
Fig.$\left(\ref{fig:Schmidt-numbermVs. Amplitude}\right)$. In the
extreme cases of either vanishing or infinite product $\sigma_{p}l$
the photons are maximally entangled $\kappa\rightarrow\infty$, and
the corresponding amplitude is $\Phi^{\left(\infty\right)}\left(\boldsymbol{q}_{s},\boldsymbol{q}_{i}\right)\propto\delta\left(\boldsymbol{q}_{s}\pm\boldsymbol{q}_{i}\right)$
as depicted in Fig.$\left(\ref{fig:Schmidt-numbermVs. Amplitude}\right)$.
We denote by $\boldsymbol{\rho}_{s/i}$  the real-space transverse
plane coordinate, conjugate to $\boldsymbol{q}_{s/i}$ . The 
real-space amplitude has two limiting cases, when $\sigma_{p}l\rightarrow0$
$\Phi\left(\boldsymbol{\rho}_{s},\boldsymbol{\rho}_{i}\right)=\Phi_{\ll}^{\left(\infty\right)}\left(\boldsymbol{\rho}_{s},\boldsymbol{\rho}_{i}\right)=\Phi_{0}\delta\left(\boldsymbol{\rho}_{s}-\boldsymbol{\rho}_{i}\right)$.
This amplitude maps the image plane explored by the signal photon
directly into the idler's detector. The opposite limiting case $\sigma_{p}L\rightarrow\infty$
is given by the amplitude $\Phi\left(\boldsymbol{\rho}_{s},\boldsymbol{\rho}_{i}\right)=\Phi_{\gg}^{\left(\infty\right)}\left(\boldsymbol{\rho}_{s},\boldsymbol{\rho}_{i}\right)=\Phi_{0}\delta\left(\boldsymbol{\rho}_{s}+\boldsymbol{\rho}_{i}\right)$.
This amplitude maps the sample plane monitored by the signal photon
$\boldsymbol{\rho}_{s}\rightarrow-\boldsymbol{\rho}_{i}$ which results
in the mirror image. We use the abbreviated notation whereby $\bar{\boldsymbol{\rho}_{i}}$
denotes the mapping from the sample to the detector plane with the
corresponding sign.\textcolor{blue}{{} }

\noindent 
\textcolor{blue}{{} }

\section{The reduced idler density matrix in the Schmidt basis \label{sec:Reduced-density-matrix-1}}

\noindent The reduced density matrix of the idler reveals the role
of quantum correlations in the proposed detection measurement scheme {[}Fig.$\text{\ensuremath{\left(\ref{The setup}\right)}}${]}.
The joint light-matter density matrix in the interaction picture is given
by,

\noindent 
\begin{equation}
\rho_{\mu\phi}^{int}\left(t\right)={\cal T}e^{-i\int d\tau{\cal H}_{I,-}\left(\tau\right)}\rho_{\mu}\otimes\rho_{\phi},
\end{equation}

\noindent where ${\cal T}$ represents super-operator time ordering
and the\emph{ off-resonance} radiation/matter coupling is ${\cal H}_{I}=\int d\boldsymbol{r}\sigma\left(\boldsymbol{r},t\right)\boldsymbol{A}^{2}\left(\boldsymbol{r},t\right)$
with the vector field $\boldsymbol{A}\left(\boldsymbol{r},t\right)=-\nicefrac{\dot{\boldsymbol{E}}\left(\boldsymbol{r},t\right)}{c}$.
The subscript $\left(-\right)$ on a Hilbert space operators represents
the commutator ${\cal O}_{-}\equiv\left[{\cal O},\cdot\right]$. The
electric field is given by $\boldsymbol{E}\left(\boldsymbol{r},t\right)=\sum_{k}\boldsymbol{E}_{k}^{\left(+\right)}\left(\boldsymbol{r},t\right)+\boldsymbol{E}_{k}^{\left(-\right)}\left(\boldsymbol{r},t\right)$
such that,

{\small{}
\begin{align}
\boldsymbol{E}_{k}^{\left(+\right)}\left(\boldsymbol{r},t\right)=\left(\boldsymbol{E}_{k}^{\left(-\right)}\left(\boldsymbol{r},t\right)\right)^{\dagger} & =\sqrt{\frac{2\pi\hbar\omega_{\boldsymbol{k}}}{V_{\boldsymbol{k}}}}\sum_{\nu}\epsilon_{\boldsymbol{k}}^{\left(\nu\right)}a_{\boldsymbol{k},\nu}e^{i\boldsymbol{k}\cdot\boldsymbol{r}-i\omega_{\boldsymbol{k}}t},
\end{align}
}{\small\par}

\noindent $\mu$ stands for the matter's degrees of freedom while
$\phi$ represents the field's degrees of freedom. For a weak field,
one can expand the evolution of the density matrix in powers of the
field which correspond to number of light-matter interactions. To
first nontrivial order, a single interaction from the left or the
right of the joint space density matrix corresponds to a change in
the coherence in the field subspace $\rho_{\phi}=\text{tr}_{\mu}\rho_{\mu\phi}$.
The radiation field records no photon exchange due to a single interaction
with the matter, merely a change in its phase. When the initial state
of the field contains an interesting internal structure such as quantum
correlations arising from entanglement, the initial reduced density
matrix $\rho_{\phi_{i}}=\text{tr}_{\mu\phi_{s}}\rho_{\mu,\phi_{si}}$
obtained by tracing over the signal beam is given by,

\noindent 
\begin{equation}
\rho_{\phi_{i}}\left(0\right)=\sum_{n,i,i'}\lambda_{n}v_{n}^{*}\left(\boldsymbol{k}_{i}\right)v_{n}\left(\boldsymbol{k}_{i}'\right)\vert\boldsymbol{1}_{i}\rangle\langle\boldsymbol{1}_{i'}\vert,\label{Density matrix before}
\end{equation}

\noindent which is diagonal in the idler subspace in the Schmidt basis.
When the signal interacts with an external matter degree of freedom,
the idler reduced density matrix is no longer diagonal. 
In the small diffraction angle limit it
is given by (see appendix 1 of the SI), 

\noindent 
\begin{equation}
\rho_{\phi_{i}}^{\left(1\right)}=\sum_{n,m,i,i'}{\cal P}_{nm}v_{n}^{*}\left(\boldsymbol{k}_{i}\right)v_{m}\left(\boldsymbol{k}_{i}'\right)\vert\boldsymbol{1}_{i}\rangle\langle\boldsymbol{1}_{i'}\vert+h.c,\label{eq: Density matrix after}
\end{equation}

\noindent where ${\cal P}_{nm}=i\beta_{nm}^{\left(1\right)}\sqrt{\lambda_{n}\lambda_{m}}$,
and,

\begin{equation}
\beta_{nm}^{\left(1\right)}=\int d\boldsymbol{r\:}u_{n}\left(\boldsymbol{r}\right)\sigma\left(\boldsymbol{r}\right)u_{m}^{*}\left(\boldsymbol{r}\right)\label{eq: Beta(1)}
\end{equation}

\noindent are the projections of matter quantities on the chosen Schmidt
basis. Our setup allows to probe the induced coherence of the field
due to its interaction with matter. 

\begin{figure}[h]
	\begin{centering}
		\includegraphics[viewport=190bp 510bp 422bp 745bp,clip]{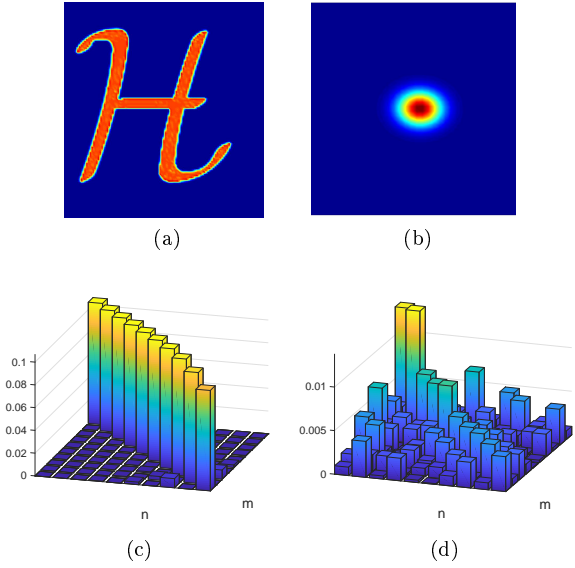}
		\par\end{centering}
	\caption{\label{Reduced density matrix figures} \emph{The reduced idler density-matrix
			 in the Schmidt basis.} $\left(a\right)$ The projected
		object. $\left(b\right)$ The 'spot-size' corresponding to the $HG_{00}$
		mode. $\left(c\right)$ The idler's reduced density matrix before
		the interaction with the object presented in Hermite-Gauss basis modes,
		given by Eq.$\text{\ensuremath{\left(\ref{Density matrix before}\right)}}$.
		$\left(d\right)$ The change in the reduced density matrix of the
		idler due to the interaction with the object given by Eq. $\left(\text{\ref{eq: Density matrix after}}\right)$}
	
\end{figure}

\begin{figure}[h]
	\begin{centering}
		\includegraphics[viewport=140bp 260bp 482bp 540bp,clip,scale=0.73]{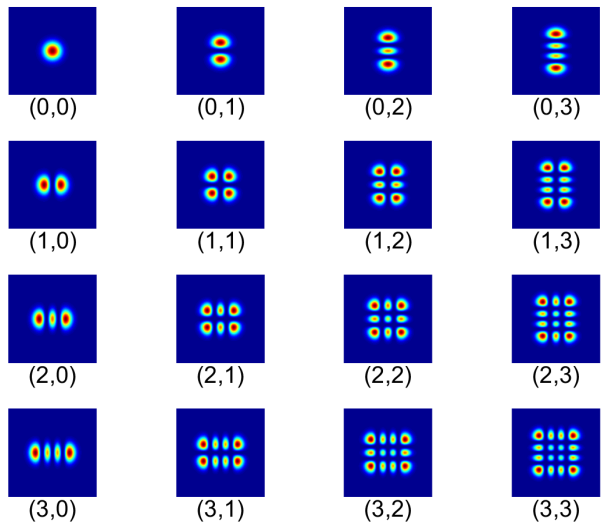}
		\par\end{centering}
	\caption{\emph{Hermite-Gaussian modes.} Modes are labeled by two indices, each
		representing one dimension in the transverse plane. \label{fig:Hermite-Gaussian-modes visualization}}
\end{figure}

Fig. $\left(\text{\ref{Reduced density matrix figures}}\right)$ displays
the induced Schmidt-space coherence of the reduced density matrix
of idler (the non-interacting photon) due to the interaction of its
twin (signal) with an object. We have chosen the Hermite-Gauss basis,
depicted in Fig.$\left(\text{\ref{fig:Hermite-Gaussian-modes visualization}}\right)$
for this visualization. Each mode is labeled by two indices, one for
each spatial dimension of the image. In Fig.$\left(\text{\ref{Reduced density matrix figures}}.c,d\right)$,
we have traced over the corresponding index,  resulting
in a one dimensional data set. Each coherence corresponds to a projection
of the object between two modes. Eq.$\left(\text{\ref{eq: small angel main eq}}\right)$
can be derived as the intensity expectation value calculated from
the idler's reduced density matrix given in Eq.$\left(\text{\ref{eq: Density matrix after}}\right)$.

\section{Far-field diffraction}

We next turn to far field diffraction with arbitrary scattering directions.
While the incoming field is understood to be paraxial, the scattered
field is not. The coincidence image in the far-field yields a similar
expression to the one calculated from the reduced density matrix in
Eq.$\left(\text{\ref{eq: Density matrix after}}\right)$ with an additional
spatial phase factor characteristic to far-field diffraction. Using
Eq.$\left(\ref{wave-function - general term}\right)$ for the setup
described in Fig.$\left(\ref{The setup}\right)$,  the coincidence
image is given by the intensity-intensity correlation function (see SI),

\begin{align}
S\left[\bar{\boldsymbol{\rho}}_{i}\right] & =\int d\boldsymbol{X}_{s}d\boldsymbol{X}_{i}G_{s}\left(\boldsymbol{X}_{s},\bar{\boldsymbol{X}_{s}}\right)G_{i}\left(\boldsymbol{X}_{i},\bar{\boldsymbol{X}_{i}}\right)\label{eq: intensity-intensity definition}\\
 & \times\left\langle {\cal T}\hat{I}_{s}\left(\boldsymbol{r}_{s},t_{s}\right)\hat{I}_{i}\left(\boldsymbol{r}_{i},t_{i}\right){\cal U}_{I}\left(t\right)\right\rangle ,\nonumber 
\end{align}

\noindent here $\hat{I}_{m}\left(\boldsymbol{r}_{m},t_{m}\right)\equiv\hat{\boldsymbol{E}}_{m,R}^{\left(-\right)}\left(\boldsymbol{r}_{m},t_{m}\right)\cdot\hat{\boldsymbol{E}}_{m,L}^{\left(+\right)}\left(\boldsymbol{r}_{m},t_{m}\right)$
are field intensity operators and $m=\left(s,i\right)$. The gating functions
$G_{m}$ represent the details of the measurement process \cite{Glauber2007,Roslyak2009}.
Eq.$\left(\text{\ref{eq: intensity-intensity definition}}\right)$ can be calculated straight from the reduced density matrix of
the idler, despite the fact that it includes the signal's intensity
operator. The reason stems from the fact that the intensity
expectation value monitors the single photon space. The partial trace
over a singly occupied signal state results in the same operation.
Estimating this expression includes a 10 field operator correlation
function which are shown explicitly in Eq(4) of appendix (2).
In the far-field, upon rotational averaging we obtain (see appendix
2 of the SI),

\begin{align}
{\cal S} & \left[\bar{\boldsymbol{\rho}}_{i}\right]\propto\mathfrak{Re}\int d\omega_{s}\mathscr{E}\left[\omega_{s}\right]\int d\boldsymbol{\rho}_{s}\Phi\left(\boldsymbol{\rho}_{s},\boldsymbol{\bar{\rho}}_{i}\right)\times\nonumber \\
& \int d\boldsymbol{\rho}^{'}\Phi\left(\boldsymbol{\rho}^{'},\bar{\boldsymbol{\rho}}_{i}\right)\boldsymbol{\boldsymbol{\sigma}}\left(\boldsymbol{\rho}^{'}\right)e^{-i\boldsymbol{\boldsymbol{Q}_{s}}\cdot\boldsymbol{\rho}^{'}}.\label{eq:Signal with general Phi-1-1}
\end{align}

\noindent Here $\boldsymbol{Q}_{s}=\frac{\omega_{s}}{c}\hat{\rho}_{s}$
is the diffraction wavector, $\mathscr{E}\left[\omega_{s}\right]=\int d\omega_{i}G\left(\omega_{s}\right)G\left(\omega_{i}\right)\left|A\left(\omega_{s}+\omega_{i}\right)\right|^{2}$
is a functional of the frequency, ${\cal S}=-\left(S-S_{0}\right)$
is the image with the noninteracting-uniform background $\left(S_{0}\right)$
subtracted, and $\bar{\boldsymbol{\rho}}_{i}$ is the mapping coordinate
onto the detector plane with the corresponding sign. $\boldsymbol{\boldsymbol{\sigma}}\left(\boldsymbol{\rho}\right)\equiv\sum_{\alpha;a,b}\left\langle a\vert\hat{\sigma}\left(\boldsymbol{\rho}-\boldsymbol{\rho}_{\alpha}\right)\vert b\right\rangle $
denotes a matrix element of the charge-density operator, traced over
the longitudinal axis, with respect to the eigenstates $\left\{ a,b\right\} $
and $\boldsymbol{\rho}_{\alpha}$ are positions of particles in the
sample. The matter can be prepared initially in a superposition state.
Substituting the Schmidt decomposition {[}Eq.$\left(\ref{eq:Schmidt decomposition}\right)${]}
into Eq.$\left(\ref{eq:Signal with general Phi-1-1}\right)$ gives,

\noindent 
\begin{align}
{\cal S} & \left[\bar{\boldsymbol{\rho}}_{i}\right]\propto\mathfrak{Re}\int d\omega_{s}\mathscr{E}\left[\omega_{s}\right]d\boldsymbol{\rho}_{s}\sum_{nm}^{\infty}\sqrt{\lambda_{n}\lambda_{m}}u_{n}\left(\boldsymbol{\rho}_{s}\right)v_{n}^{*}\left(\boldsymbol{\bar{\rho}}_{i}\right)\times\nonumber \\
 & v_{m}\left(\boldsymbol{\bar{\rho}}_{i}\right)\int d\boldsymbol{\rho}^{'}u_{m}^{*}\left(\boldsymbol{\rho}^{'}\right)\boldsymbol{\boldsymbol{\sigma}}\left(\boldsymbol{\rho}^{'}\right)e^{-i\boldsymbol{Q}_{s}\cdot\boldsymbol{\rho}^{'}}.\label{eq:Signal Schmidt decomposition-1-1}
\end{align}

\noindent This shows a smooth transition from momentum to real space
imaging. For low Schmidt modes that do not vary appreciably across
the charge density scale, the last term yields $\boldsymbol{\boldsymbol{\sigma}}\left(\boldsymbol{Q}_{s}\right)\approx\int d\boldsymbol{\rho}^{'}u_{m}^{*}\left(\boldsymbol{\rho}^{'}\right)\boldsymbol{\boldsymbol{\sigma}}\left(\boldsymbol{\rho}^{'}\right)e^{-i\boldsymbol{Q}_{s}\cdot\boldsymbol{\rho}^{'}}$.
Consequently, when the Schmidt modes do not vary on the lengthscale
of the charge density up to high order, the Fourier decomposition
of the charge density is projected on $u_{n}$ and reweights the corresponding
idler modes. The resulting image given by spatial scanning of the
idler is the Fourier transform of the charge density projected on
the relevant idler mode. Alternately, when the Schmidt modes vary along the charge
density, the exact expression for the far field diffraction image
is given by,

\noindent 
\begin{align}
{\cal S}\left[\bar{\boldsymbol{\rho}}_{i}\right] & \propto\mathfrak{Re}\sum_{nm}^{\infty}\gamma_{nm}\sqrt{\lambda_{n}\lambda_{m}}v_{n}^{*}\left(\boldsymbol{\bar{\rho}}_{i}\right)v_{m}\left(\boldsymbol{\bar{\rho}}_{i}\right)\\
\gamma_{nm} & =\sum_{k}\beta_{km}^{\left(1\right)}\int d\boldsymbol{\rho}_{s}d\omega_{s}\mathscr{E}\left[\omega_{s}\right]u_{n}\left(\boldsymbol{\rho}_{s}\right)u_{k}^{*}\left(\boldsymbol{Q}_{s}\right),
\end{align}

\noindent where $\beta_{nm}^{\left(1\right)}$ was defined in Eq.$\left(\text{\ref{eq: Beta(1)}}\right)$.
From the definition of $\boldsymbol{Q}_{s}$ it is evident that its
angular component of $u_{k}$ is identical to the corresponding in
$u_{n}$ and therefore $\gamma_{nm}$ is composed of summation over
modes with the same angular momentum in the LG basis set. 

It is also possible to calculate the real-space image of the charge density
when the signal is frequency dispersed. Assuming for simplicity perfect
quantum correlations between the signal and idler we obtain,

\begin{equation}
{\cal S}\left[\bar{\boldsymbol{\rho}}_{i},\bar{\omega}_{s}\right]\propto\mathfrak{Re}\sigma\left(\boldsymbol{\bar{\rho}}_{i}\right)e^{-i\frac{\bar{\omega}_{s}}{c}\bar{\rho}_{i}}.
\end{equation}

\noindent This image is phase dependent and unlike diffraction of classical light, allows to transform
freely between momentum and real-space. The phase-dependent Fourier image
in this limit is also given by resolving the signal photon with respect
to the frequency $\bar{\omega}_{s}$ as well (see SI). 

\begin{figure}[h]
	\begin{centering}
		\includegraphics[viewport=202bp 380bp 422bp 742bp,clip,scale=1.1]{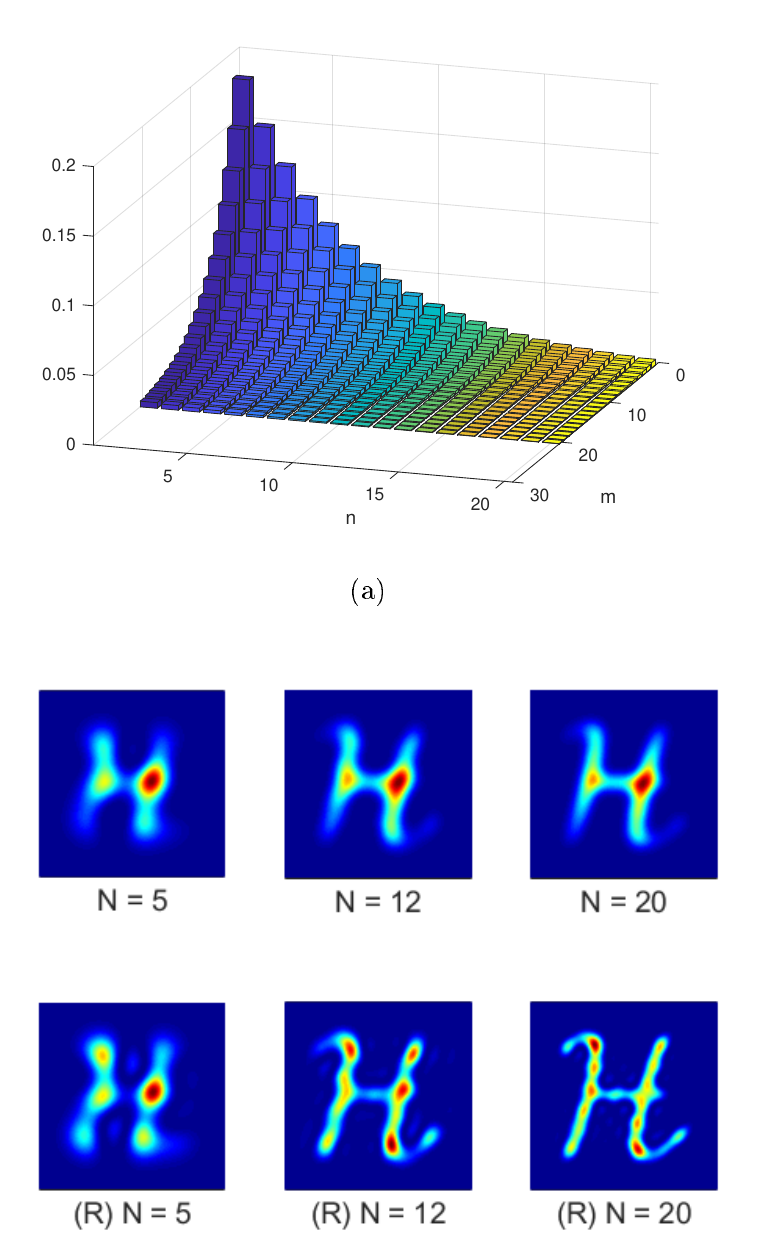}
		\par\end{centering}
	\caption{\label{fig:Weighted-recombination}Weighted recombination of the truncated
		sum in Eq.$\left(\text{\ref{eq:projection}}\right)$, using HG basis
		with $\sigma_{p}l=0.07$, corresponding to $\kappa\approx14$. $\left(a\right)$
		Schmidt weights of the entangled light source. $\left(b\right)$ First
		order image. Recombination using the original weight of each mode
		(upper row), with respect to the $N$ first modes. This corresponds
		to straightforward imaging with the given parameters. The lower row
		shows the reweighted-flattened Schmidt spectrum recombination that
		corresponds to the $N$ first modes, marked with (R). $\left(c\right)$
		The real part of the image $I\left(r,\phi\right)$ with added spatial
		phase $\left|I\left(\rho,\phi\right)\right|\exp\left[-i\frac{2\pi}{L/3}\rho\right]$.
		$\left(d\right)$ Reweighted truncated sum diffraction image given
		by Eq.$\left(\text{\ref{eq:projection}}\right)$ for N=20. Recovering
		the spatial phase.}
\end{figure}

\section{Reweighted modal-contributions }

The apparent classical-like form of the coherent superposition in
the Schmidt representation, where each mode carries a distinct spatial
matter information, suggests experiments in which a single Schmidt mode
is measured at a time \cite{Straupe2011}. This bares some resembles to the coherent 
mode representation of partially coherent sources studied in \cite{Wolf1982,Vartanyants2018}. 
Moreover, it allows the reweighting
of high angular momentum modes available experimentally \cite{Fickler2016},
and known to have decreasing effect on the image upon naive summation.
Reweighting of truncated sums is extensively used as sharpening tool
in digital signal processing, especially in medical image enhancement
\cite{Lehmann1999}.  This approach raises questions regarding the
analysis of optimal Schmidt weights, error minimization and engineered
functional decrease of weights as done in theory of sampled signals. 
The structure of the spatial information mapping from the signal to
the idler takes a simpler form for small scattering angles. When we
examine the first and second order contributions due to a single charge
distribution, the resulting image of a truncated sum composed of the
first $N$ modes is given by,

\begin{align}
{\cal S}_{N}^{\left(p\right)}\left[\bar{\boldsymbol{\rho}}_{i}\right] & \propto\mathfrak{Re}\sum_{nm=0}^{N}\sqrt{\lambda_{n}\lambda_{m}}\beta_{nm}^{\left(p\right)}v_{n}^{*}\left(\boldsymbol{\bar{\rho}}_{i}\right)v_{m}\left(\boldsymbol{\bar{\rho}}_{i}\right),\label{eq:projection}
\end{align}

\noindent where $\beta_{nm}^{\left(2\right)}=\int d\boldsymbol{r}\:u_{n}\left(\boldsymbol{r}\right)\left|\sigma\left(\boldsymbol{r}\right)\right|^{2}u_{m}^{*}\left(\boldsymbol{r}\right),$
is a scattering coefficient between Schmidt modes which resembles
the expressions used in previous two-photon imaging techniques
\cite{Shapiro2008,Boyd2008,Walborn2010}. $\beta_{nm}^{\left(1\right)}$
defined in Eq.$\left(\text{\ref{eq: Beta(1)}}\right)$, holds phase
information of the studied object and have no classical counterpart.
Its momentum space representation reads, 

\noindent 
\begin{equation}
\beta_{nm}^{\left(1\right)}=\sum_{\boldsymbol{k}_{s},\boldsymbol{k}_{d}}u_{n}\left(\boldsymbol{k}_{s}\right)\sigma\left(\boldsymbol{k}_{s}-\boldsymbol{k}_{d}\right)u_{m}^{*}\left(\boldsymbol{k}_{d}\right)
\end{equation}

\noindent where $d$ stands for a detected mode initially in a vacuum
state. This shows more clearly the physical role played by the charge
density in the coupling of different Schmidt modes.

\noindent Fig.$\left(\ref{fig:Weighted-recombination}a\right)$ presents
the Schmidt spectrum for a beam characterized by $\sigma_{p}l=0.07$
which yields $\kappa\approx14$. Fig.$\left(\ref{fig:Weighted-recombination}b\right)$
illustrates the improvement of the acquired image due to resummation
of the Hermite-Gauss modes of the object decomposed in Fig. $\left(\text{\ref{Reduced density matrix figures}}\right)$.
By using Eq.$\left(\text{\ref{eq:projection}}\right)$ with flattened
Schmidt spectrum we demonstrate the enhancement of fine features of
the diffracted image. Phase measurement is demonstrated in Fig.$\left(\ref{fig:Weighted-recombination}c,d\right)$.

\section{Discussion}

The scattered quantum light from matter carries phase information at odd
orders in the charge distribution $\sigma\left(\boldsymbol{q}\right)$
the light-matter interaction. To first order, the change in the quantum
state of the field due to a single interaction is imprinted in the
phase of the photons, which is detectable. However, no photon is generated
in this order. Homodyne diffraction of classical sources results in
even correlation functions of the charge density. We have provided
a complete description of the charge distribution resulting from nonvanishing
odd orders of the radiation-matter interaction. The detected image
is sensitive to the degree of entanglement. High resolution is achieved
in the limits of infinite or vanishing $\sigma_{p}l$, which are hard
to realize. For a long nonlinear crystal, the phase matching factor
is more dominant and strong beam divergence is required to generate
strong quantum correlations. This limit is not compatible with the
paraxial approximation for the amplitude and requires further study.
In the short crystal limit the amplitude acquires the angular spectrum
of the pump and the resolution is limited by the crystal length and
low beam divergence. 

We have demonstrated that coincidence diffraction measurements of
entangled photons with quantum detection, can also achieve enhanced
imaging resolution. Eq.$\left(\ref{eq:projection}\right)$ provides
an intuitive picture for the information transfer from
the signal to the idler beams. By reweighting the spatial modes that
span the measured image, one can refine the matter information. High
angular momentum states of light have been recently demonstrated experimentally
with quantum numbers above $\sim10^{4}$ \cite{Fickler2016}. It is
of cardinal practical importance to quantify the natural cutoff of
high topologically charged modes in order to discuss sub-wavelength
resolution. Reweighting the Schmidt modes distribution is motivated
by the closure relations $\sum_{n}\:u_{n}\left(\boldsymbol{q}\right)v_{n}\left(\boldsymbol{q}^{'}\right)=\delta^{\left(2\right)}\left(\boldsymbol{q}-\boldsymbol{q}^{'}\right)$.
This suggests that equal contribution of modes converges into a delta
distribution of the two photon amplitude, perfectly transferring the
spatial information between the photons. Finding optimal weights is
a challenge for future studies. Signal acquisition optimization techniques
used in sampling theory, avoiding high frequency quantization noise
can be considered as well \cite{Lehmann1999}. 

The imaging of single localized biological molecules has been a major
driving force for building free electron X-ray lasers \cite{Chapman2011}.
Such molecules are complex, fragile, and typically have multiple timescale
dynamics. One strategy is to use a fresh sample in each iteration,
assuming a destructive measurement. Ultra-short X-ray pulses have
been proposed to reduce damage \cite{Neutzo2000}. Entangled hard
X-ray photons have been generated by parametric down conversion using
a diamond crystal \cite{Shwartz2012}. Avoiding damage of such complexes
by using weak fields, allows to follow the evolution of initially
perturbed charge densities. Linear diffraction scales as $\propto I_{p}^{\nicefrac{1}{2}}$
with the signal photons that interact with the sample while the over
all coincidence image scales as $\propto I_{p}^{\nicefrac{3}{2}}$.
Using diffraction of entangled photons from charge distributions initially
prepared by ultrafast pulses, results in imaging of their real-space
dynamics and provides a fascinating topic for future study.

\textbf{\emph{Acknowledgments.}} The support of the Chemical Sciences, 
Geosciences, and Biosciences Division, Office of Basic Energy 
Sciences, Office of Science, U.S. Department of Energy is gratefully acknowledged. Collaborative visits of K.D. to UCI were supported by Award No. DEFG02-04ER15571 ,and. S.M was supported by   Award DESC0019484. S.A   fellowship was
supported by  the National Science Foundation (Grant No. CHE-1663822) .  We  also wish to thank Noa Asban for the graphical illustrations.

\bibliography{Quantum_imaging_main}

\end{document}